\documentclass[twocolumn,prb]{revtex4}
\usepackage{amsfonts}
\usepackage[T1]{fontenc}
\usepackage{amsmath,amsbsy,amssymb,graphicx}
\usepackage{times}

\def\beginABC{\begin{subequations}}
\def\endABC{\end{subequations}}
\let\mathbf=\boldsymbol
\def\Section#1{\bigskip\noindent\textbf{\large#1}\par\noindent}
\let\section=\Section

\begin{document}

\title{{\Large Electrically Tunable Conductance and Edge Modes}\\
{\Large in Topological Crystalline Insulator Thin Films:}\\
{\Large Tight-Binding Model Analysis}}
\author{Motohiko Ezawa}
\affiliation{Department of Applied Physics, University of Tokyo, Hongo 7-3-1, 113-8656,
Japan }

\begin{abstract}
We propose a minimal tight-binding model for thin films made of topological
crystalline insulator (TCI) on the basis of the mirror and discrete
rotational symmetries. The basic term consists of the spin-orbit interaction
describing a Weyl semimetal, where gapless Dirac cones emerge at all the
high symmetry points in the momentum space. We then introduce the mass term
providing gaps to Dirac cones at our disposal. They simulate the thin films
made of the [001], [111] and [110] TCI surfaces. TCI thin films are
topological insulators protected by the mirror symmetry. We analyze the
mirror-Chern number, the edge modes and the conductance by breaking the
mirror symmetry with the use of electric field. We propose a multi-digit
topological field-effect transistor by applying electric field independently
to the right and left edges of a nanoribbon. Our results will open a new way
to topological electronics.
\end{abstract}

\maketitle


\address{{\normalsize Department of Applied Physics, University of Tokyo, Hongo
7-3-1, 113-8656, Japan }}

Topological insulator is one of the most fascinating concept found in this
decade\cite{Hasan,Qi}. Recent flourish of the study of topological insulator
is based on the finding of the time-reversal invariant topological
insulator\cite{KaneMele,FuKaneMele,TFT}. Very recently, a new class of topological insulator, topological
crystalline insulator (TCI), attracts much attention\cite%
{Fu,Hsieh,Ando,Xu,Dz,Tanaka,Polly,LiuFu,FangBer,LiuFu04,Fang,Okada,Safaei,Wojek,Wang}%
. The best example is given by Pb$_{x}$Sn$_{1-x}$Te, which has been found to
be a TCI experimentally\cite{Ando,Xu,Dz}. There are three types of the
surface states, the [001], [111] and [110] surfaces, which have discrete
rotation symmetries $C_{N}$ with $N=4$, $6$ and $2$, respectively. Gapless
Dirac cones emerge inevitably on the surface of a topological insulator.
Indeed, it has been experimentally observed that there are gapless Dirac
cones at the $X$ and $Y$ points in the [001] surface\cite{Ando,Xu,Dz}, and
at the $\Gamma $ and three $M$ points in the [111] surface\cite{Tanaka,Polly}%
. Theoretical studies have been performed with the use of first-principle
calculation and low-energy effective Dirac theory.

Although there are yet no experimental measurements, theoretical studies 
\cite{LiuFu,FangBer} have been presented also on the thin film made of a
TCI. It is characterized by the same discrete rotation symmetry $C_{N}$, and
additionally by the mirror symmetry about the 2D plane. When the film
is thin enough, a gap opening occurs due to hybridization between the front
and back surfaces, and it turns the system into a topological insulator.
Now, electronics is based on circuits in two dimensions. It is important and
urgent to make a further investigation of a TCI thin film, since it may well
be a good candidate for designing nanodevices in topological electronics.

The TCI thin film is a topological insulator protected by the
mirror-symmetry. Accordingly the topological number is the mirror-Chern
number\cite{TeoMirror,Takahashi}. A prominent feature is that we can break
the mirror symmetry simply by applying external electric field. This is highly
contrasted to the case of the time-reversal invariant topological insulator,
where the time-reversal breaking should be caused by magnetic field or
exchange field induced by ferromagnet. Magnetic field and exchange field are
hard to control precisely. On the other hand, a precise control of the
electric field is possible within current technique.

The aim of this work is to explore the physics of the TCI thin film by
constructing a minimal tight-binding model based on the discrete rotation
symmetry $C_{N}$ and the mirror symmetry about the 2D plane. The
tight-binding model is essential to make a deeper understanding of the
system, which is not attained by first-principle calculation and low-energy
effective Dirac theory. For instance, according to the low-energy theory 
\cite{LiuFu}, the [001] thin film made of Pb$_{x}$Sn$_{1-x}$Te has two Dirac
cones at the $X$ and $Y$ points, where the chirality is identical. However,
the Nielsen-Ninomiya theorem\cite{Nielsen} dictates that the total chirality
is zero. Consequently, there must be two additional Dirac cones with the
chirality opposite to that of the $X$ and $Y$ points. In our tight-binding
model these two Dirac cones emerge at the $\Gamma $ and $M$ points, though
they are removed from the low-energy spectrum. The tight-binding model is
also useful to analyze the edge states, which transport the ordinary
electric current reflecting the topological properties of the thin film.

We start with a system where the spin-orbit interaction (SOI) dominates the
transfer term. We demonstrate that such a system is a Weyl semimetal consisting of 
multiple Dirac cones at
all the high symmetric points. We next introduce mass terms to give gaps to
Dirac cones at our disposal. The system may well describe the TCI thin film
made of Pb$_{x}$Sn$_{1-x}$Te with an appropriate choice of phenomenological
mass parameters. Then we break the mirror symmetry by introducing a
perpendicular electric field $E_{z}$. The conductance is calculated in the presence of $E_{z}$. The conductance is switched off by the electric field. 
Namely, it acts as a topological field-effect transistor \cite%
{EzawaAPL}. By attaching two independent gates to the sample, we can
separately control the right and left edge states. The conductance can be 0,
1 and 2, which forms a multi-digit topological field-effect transistor,
where the conductance is quantized and topologically protected. Our results
open a new way to electric-field controllable topological electronics.

\section{Main Results}

Our main results consist of a phenomenological construction of minimal
tight-binding Hamiltonians for the TCI thin film, and the analysis of
electrically controllable conductance and edge modes of a nanoribbon in view
of the bulk-edge correspondence.

\textbf{Tight-binding Hamiltonians.} The SOI plays a key role in the physics
of topological insulators. We consider a model where the SOI dominates the
system. A simplest example would be the Rashba SOI,%
\begin{equation}
H_{\text{SO}}=i\sum_{\ell =1}\lambda _{\ell }\sum_{\left\langle
i,j\right\rangle }c_{i}^{\dagger }[\mathbf{\sigma }\times \mathbf{d}%
_{ij}^{\ell }]c_{j}.  \label{HamilR}
\end{equation}%
Alternatively we may think of%
\begin{equation}
H_{\text{SO}}=i\sum_{\ell =1}\lambda _{\ell }\sum_{\left\langle
i,j\right\rangle }c_{i}^{\dagger }[\mathbf{\sigma }\cdot \mathbf{d}%
_{ij}^{\ell }]c_{j},  \label{Hamil}
\end{equation}%
or even take a sum of them. Here, $\mathbf{\sigma }=(\sigma _{x},\sigma
_{y},\sigma _{z})$ represents the Pauli matrix for the spin, and $\mathbf{d}%
_{ij}^{\ell }=\mathbf{r}_{i}-\mathbf{r}_{j}$ connects a pair of the $\ell $%
-th nearest neighbor sites $i$ and $j$ in the lattice with $\lambda _{\ell }$
the coupling strength. As is easily shown, the results based on the
Hamiltonian (\ref{HamilR}) are obtained from those on the Hamiltonian (\ref%
{Hamil}) simply with the replacement of a set of momenta $(k_{x},k_{y})$ by $%
(k_{y},-k_{x})$. Furthermore, the low-energy theory derived from a
first-principle calculation\cite{LiuFu} supports the choice of (\ref{Hamil}). 
Hence we concentrate on (\ref{Hamil}) hereafter.

\begin{figure}[t]
\centerline{\includegraphics[width=0.48\textwidth]{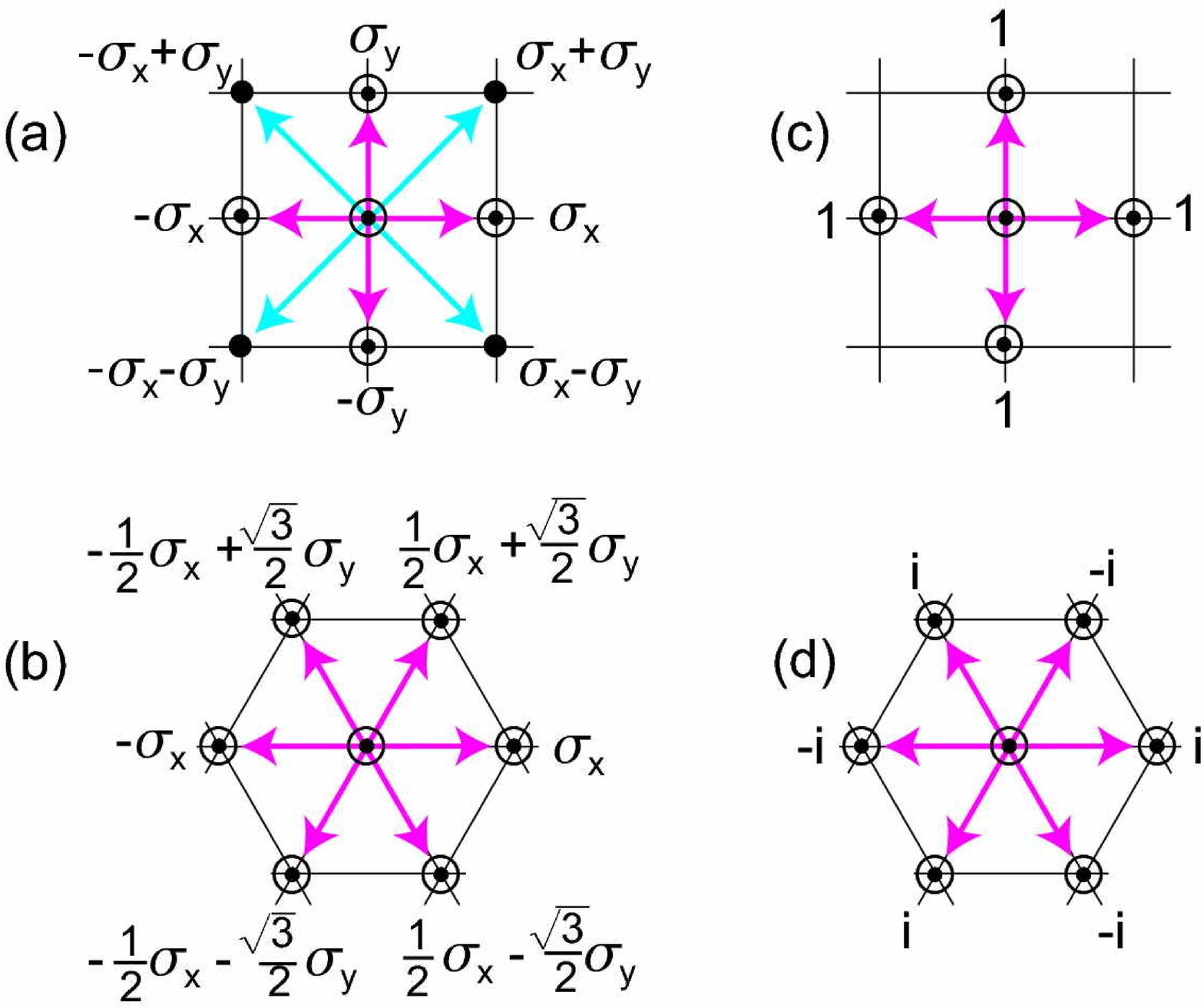}}
\caption{\textbf{Nearest and next-nearest neighbor sites in real space.} We
give the value of  $\mathbf{\protect\sigma }\cdot \mathbf{d}_{ij}^{\ell }$ for the
nearest ($\ell =1$, magenta) and next-nearest ($\ell =2$, cyan) neighbor
sites for the square lattice (a) and the triangular lattice (b) in the
Hamiltonian (\protect\ref{Hamil}). We also give the value of $\protect\nu _{ij}$
for the nearest neighbor site ($\ell =1$, magenta) for the square lattice
(c) and the triangular lattice (d) in the Hamiltonian (\protect\ref{Hamil-M}%
). }
\label{FigLattice}
\end{figure}

Let $N_{\ell }$ be the number of the $\ell $-th neighbor sites. In the
momentum representation the Hamiltonian is rewritten as%
\begin{equation}
H_{\text{SO}}=\sum_{\mathbf{k}}c^{\dagger }(\mathbf{k})\hat{H}_{\text{SO}}(%
\mathbf{k})c(\mathbf{k}),  \label{Hamil-SO}
\end{equation}%
with%
\begin{equation}
\hat{H}_{\text{SO}}(\mathbf{k})=\sum_{\ell =1}\lambda _{\ell
}\sum_{n=1}^{N_{\ell }}\mathbf{\sigma }\cdot \mathbf{d}_{n}^{\ell }e^{i%
\mathbf{d}_{n}^{\ell }\cdot \mathbf{k}},  \label{Hamil-SOO}
\end{equation}%
where $\mathbf{d}_{ij}^{\ell }$ is determined by requiring the invariance
under the discrete rotational symmetry $C_{N}$. (See the Methods with
respect to $C_{N}$.) For instance, $\mathbf{d}_{n}^{\ell }$ for the nearest
neighbor sites ($\ell =1$) is expressed as%
\begin{equation}
\mathbf{d}_{n}^{1}=|\mathbf{d}_{n}^{1}|\left( \cos \theta _{n},\sin \theta
_{n}\right) ,\quad \theta _{n}=\frac{2\pi n}{N}
\end{equation}%
for the triangular ($N=3$) and square ($N=4$) lattices: See Fig.\ref%
{FigLattice}(a) and (c). We shall soon see that this model has multiple
Dirac cones at the high symmetry points known such as the $X$, $Y$, $\Gamma $
and $M$ points in the square lattice and the $\Gamma ,K,K^{\prime
},M_{1},M_{2},M_{3}$ points in the triangular lattice. Thus the Hamiltonian (%
\ref{HamilR}) describes a Weyl semimetal.

The minimal tight-binding Hamiltonian of a TCI thin film would be a
four-band model due to the spin and pseudospin (surface) degrees of freedom.
Let $\mathbf{\tau }=(\tau _{x},\tau _{y},\tau _{z})$ be the Pauli matrix to
describe the pseudospin representing the front ($\tau _{z}=1$) and back ($%
\tau _{z}=-1$) surfaces. When the film is thin enough, the symmetric
state becomes the ground state, opening a gap to all Dirac cones due to
hybridization. We employ the Hamiltonian (\ref{HamilR}) to describe the
symmetric state. Furthermore we apply the electric field $E_{z}$ between the
two surfaces.

These effects are realized by considering the four-band effective
tight-binding Hamiltonian,%
\begin{equation}
\hat{H}=\hat{H}_{\text{SO}}\tau _{y}+\hat{H}_{m}\tau _{x}+E_{z}\tau _{z},
\label{HamilTotal}
\end{equation}%
together with $\hat{H}_{\text{SO}}$ given by (\ref{Hamil-SOO}) and $\hat{H}%
_{m}$ obtained from%
\begin{equation}
H_{m}=\sum_{\ell =0}m_{\ell }\sum_{\left\langle i,j\right\rangle }\nu
_{ij}^{\ell }c_{i}^{\dagger }c_{j},  \label{Hamil-M}
\end{equation}%
where $\nu _{ij}^{\ell }$ is a number characteristic to the lattice
structure and determined by the vector $\mathbf{d}_{ij}^{\ell }$ so as to
preserve the crystalline symmetry, and $i,j$ runs over the $\ell $-th
nearest neighbor sites. We take $\nu _{ij}^{0}=\delta _{ij}$ and show $\nu
_{ij}^{1}$ in Fig.\ref{FigLattice}(c) and (d) for the square and triangular
lattices. As we shall soon see, the gap at each Dirac point is adjusted by
choosing the mass parameters $m_{\ell }$ appropriately.

In the absence of the external electric field ($E_{z}=0$), the Hamiltonian (%
\ref{HamilTotal}) is invariant under the mirror symmetry about the 2D plane, 
\begin{equation}
MH(\mathbf{k})M^{-1}=H(\mathbf{k}),
\end{equation}%
where the mirror operator is given by 
\begin{equation}
M=-i\sigma _{z}\tau _{x}.  \label{MirrorGener}
\end{equation}%
The mirror symmetry is broken by the external electric field ($E_{z}\neq 0$)
as%
\begin{equation}
M(E_{z}\tau _{z})M^{-1}=-E_{z}\tau _{z}.  \label{MirrorE}
\end{equation}%
When the system is an insulator, the mirror-Chern charge is defined and
calculable even for $E_{z}\neq 0$ according to a general scheme\cite{SPT}.

\begin{figure}[t]
\centerline{\includegraphics[width=0.48\textwidth]{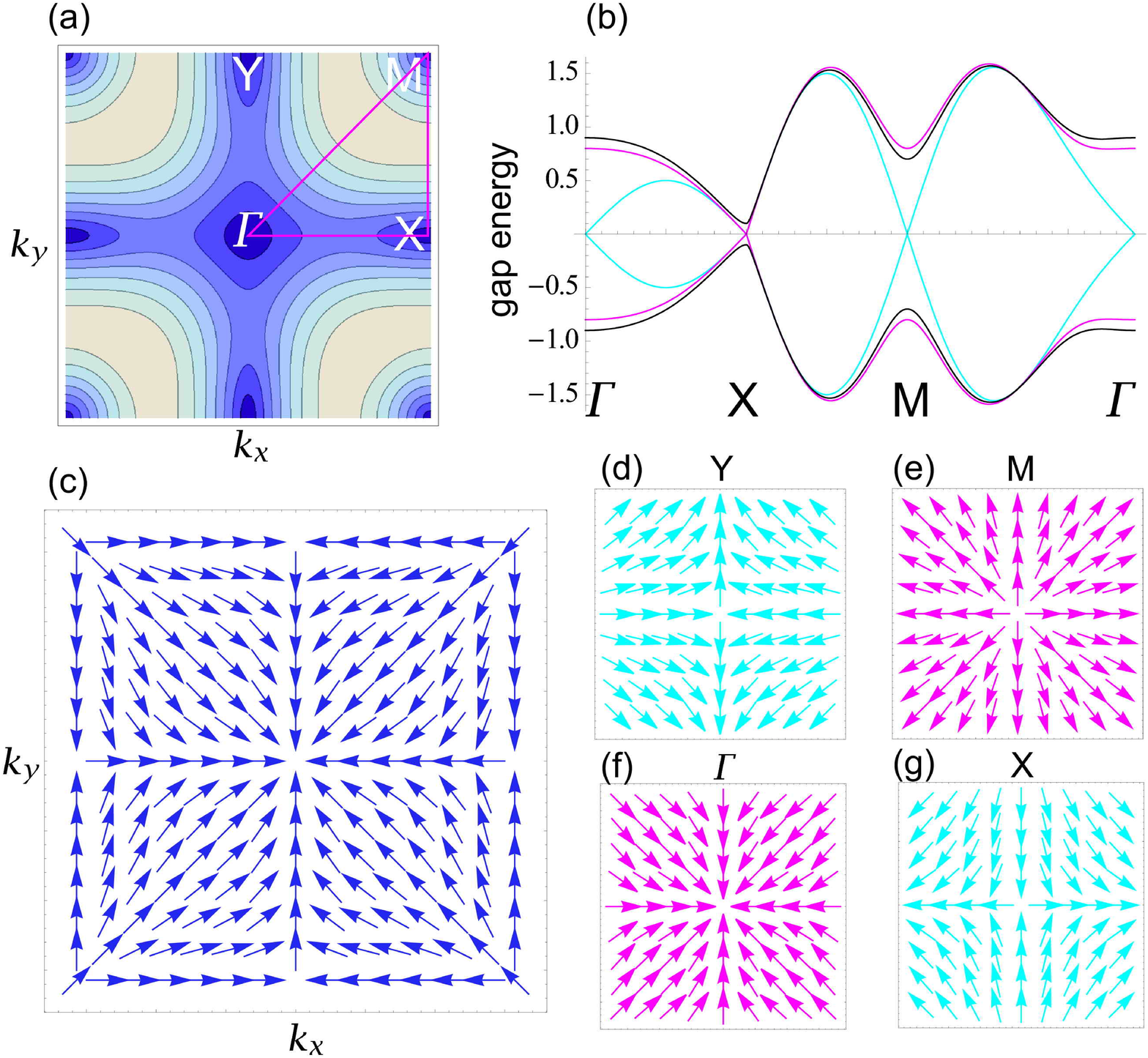}}
\caption{ \textbf{Band structure and spin direction on square lattice.} (a)
We present a contour plot of the band structure of the Hamiltonian (\protect
\ref{HamilTotal}) on the square lattice. We have set $\protect\lambda _{1}=%
\protect\lambda _{2}=0$, $m_{1}=m_{2}=0$. The energy is lower in darker
region. (b) We give the band structure along the lines shown in magenta in
(a). We have set $\protect\lambda _{1}=\protect\lambda _{2}=0$ in all three
curves. Cyan curves are with $m_{1}=m_{2}=0$. Magenta curves are with $%
m_{1}=0.4$, $m_{0}=0$. Black curves are with $m_{1}=0.4$, $m_{0}=0.1$. (c)
We show the spin direction in the thin film with the $C_{4}$ symmetry for
the whole Brillouin zone, around (d) the Y point, (e) the M point, (f) the $%
\Gamma $ point and (g) the X point. The spin structure is hedgehog-like
around the $\Gamma $ and $M$ points with the opposite directions, both
yielding the positive chirality (magenta), while it is anti-hedgehog-like
around the $X$ and $Y$ points with the opposite directions, both yielding
the negative chirality (cyan). They contribute to the mirror-Chern number.}
\label{FigSpinConfig}
\end{figure}

\textbf{Square lattice with }$C_{4}$ \textbf{symmetry.} We first consider
the square lattice with the $C_{4}$ symmetry. Let us set $E_{z}=0$. First,
taking the contributions from the nearest neighbor sites ($\ell =1)$ and the
next-nearest neighbor sites ($\ell =2$), we obtain from the Hamiltonian (\ref%
{Hamil}) as%
\begin{equation}
\hat{H}_{\text{SO}}=A_{x}\sigma _{x}+A_{y}\sigma _{y}  \label{Hamil-1}
\end{equation}%
with\beginABC\label{Axy}%
\begin{eqnarray}
A_{x} &=&\lambda _{1}\sin k_{x}+\lambda _{2}\sin k_{x}\cos k_{y},  \label{Ax}
\\
A_{y} &=&\lambda _{1}\sin k_{y}+\lambda _{2}\sin k_{y}\cos k_{x}.  \label{Ay}
\end{eqnarray}%
\endABC See the illustration in Fig.\ref{FigLattice}(a) and (b). The energy
spectrum is given by 
\begin{equation}
E=\pm \sqrt{A_{x}^{2}+A_{y}^{2}}.
\end{equation}%
There are gapless Dirac cones at the $X$, $Y$, $\Gamma $ and $M$ points, as
illustrated in Fig.\ref{FigSpinConfig}, where the band structure is shown.
The Hamiltonian describes a Weyl semimetal.

The effective low-energy Hamiltonian is given by (\ref{Hamil-1}) near each
Dirac point with%
\begin{equation}
A_{x}=v_{x}\tilde{k}_{x},\quad A_{y}=v_{y}\tilde{k}_{y},
\end{equation}%
where $v_{x}$ and $v_{y}$ are the velocities%
\begin{equation}
v_{x}=n_{x}(\lambda _{1}+n_{y}\lambda _{2}),\quad v_{y}=n_{y}(\lambda
_{1}+n_{x}\lambda _{2}),
\end{equation}%
and $\tilde{k}_{x}$ and $\tilde{k}_{y}$ are the renormalized momenta%
\begin{equation}
\tilde{k}_{x}=k_{x}+\frac{n_{x}-1}{2}\pi ,\quad \tilde{k}_{y}=k_{y}+\frac{%
n_{y}-1}{2}\pi ,
\end{equation}%
as follows from (\ref{Axy}). A set of numbers $(n_{x},n_{y})$ is $(-1,1)$
for $X$, $(1,-1)$ for $Y$, $(1,1)$ for $\Gamma $, $(-1,-1)$ for $M$. The
chirality of the Dirac cone is give by $n_{x}n_{y}$ at each point. An
anisotropy ($v_{x}\neq v_{y}$) has been introduced into the system by
introducing the nearest and next--nearest neighbor contributions ($\lambda
_{1}\neq 0,\lambda _{2}\neq 0$).

\begin{figure*}[t]
\centerline{\includegraphics[width=0.9\textwidth]{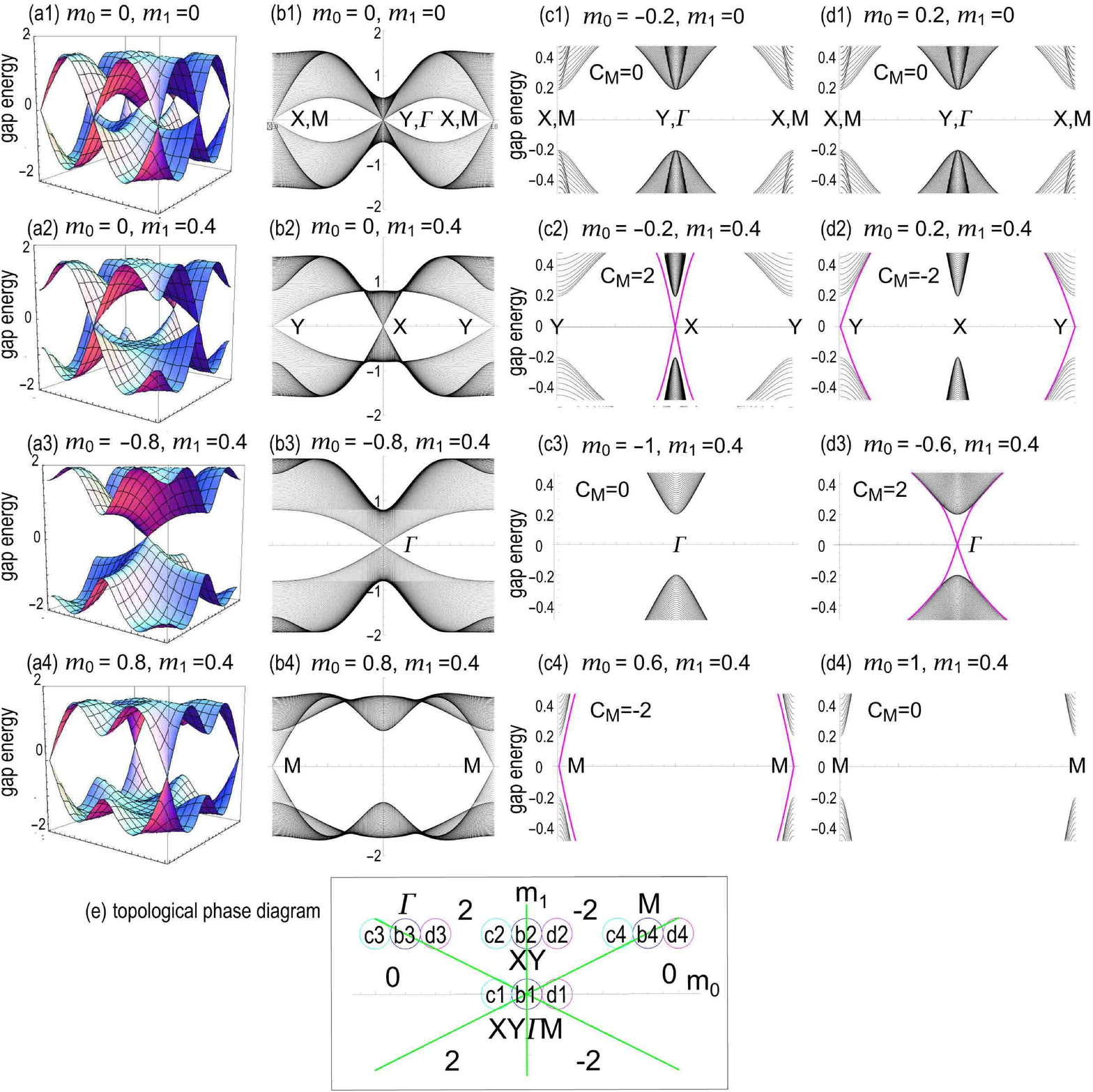}}
\caption{\textbf{Band structure and topological phase diagram for square
lattice.} (a) We show bird's eye views and (b) projected views of the bulk
band structure. (c),(d) We show the band strucure of nanoribbons, where the
gapless edge states are depicted in the magenta curves. They emerge in
topological insulators with nonzero mirror-Chern numbers. We have taken $%
\protect\lambda _{1}=1$ and $\protect\lambda _{2}=-0.5$ in all figures. The
values of $m_{0}$ and $m_{1}$ are indicated in each figure. The vertical
axis is the energy in unit of $\protect\lambda _{1}$ in all figures. The
horizontal axes are $-\protect\pi <k_{x}\leq \protect\pi $, $-\protect\pi %
<k_{y}\leq \protect\pi $ in (a), $-\protect\pi <k_{x}\leq \protect\pi $ in
(b), $-\protect\pi <k\leq \protect\pi $ in (c) and (d). (e) We present the
topological phase diagram in the $m_{0}$-$m_{1}$ plane. A green line
represents a phase boundary. The numbers $0$ and $\pm 2$ are the
mirror-Chern numbers. A circle with symbol such as c3 shows a point where
the band structure is calculated in (c3). }
\label{FigBulkBand}
\end{figure*}

\begin{figure*}[t]
\centerline{\includegraphics[width=0.8\textwidth]{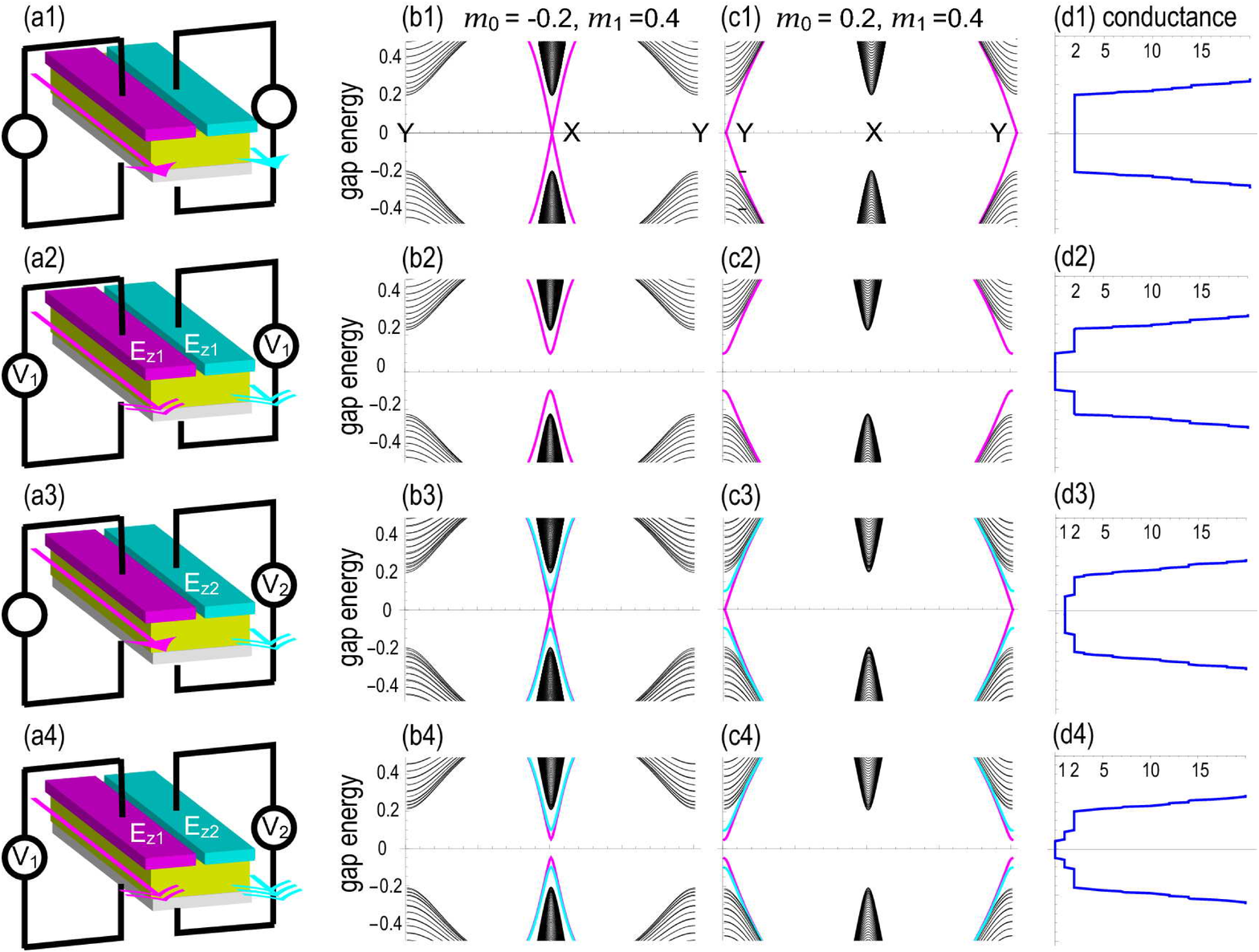}}
\caption{\textbf{Multi-digit topological transistor, edge modes and
conductance for square lattice.} (a) We apply electric fields independently to the right and
left edges of a nanoribbon. (b), (c) Gapless edge modes emerge without
electric fields since they are protected by the mirror symmetry. The gap can
be controlled by electric field independently on the right and left edges.
(d) We have calculated the conductance as a function of the gap energy. The
conductance can take quantized values $0$, $1$ and $2$ as in (d4), hence
providing us with a basic mechanism of multi-digit topological transistor.}
\label{FigEz}
\end{figure*}

We illustrate the spin direction around each Dirac point in Fig.\ref%
{FigSpinConfig}. The spin direction yields one negative chirality at the $X$
and $Y$ points apiece, while it yields one positive chirality to the $\Gamma 
$ and $M$ points apiece. The total chirality is zero over the Brillouin
zone, as required by the Nielsen-Ninomiya theorem\cite{Nielsen}.

We proceed to consider the total Hamiltonian (\ref{HamilTotal}) with $%
E_{z}=0 $, which reads 
\begin{equation}
\hat{H}=\left[ A_{x}\sigma _{x}+A_{y}\sigma _{y}\right] \tau _{y}+m\tau _{x},
\label{ThinFilm}
\end{equation}%
with (\ref{Axy}) and 
\begin{equation}
m=m_{0}+m_{1}(\cos k_{x}+\cos k_{y}).  \label{Az}
\end{equation}%
The energy spectrum is now given by 
\begin{equation}
E=\pm \sqrt{A_{x}^{2}+A_{y}^{2}+m^{2}}.
\end{equation}%
We see that $m_{1}$ opens a gap at the $\Gamma $ and $M$ points, while $%
m_{0} $ opens a gap at all Dirac points, as illustrated in Fig.\ref%
{FigSpinConfig}(b). Note that, if we set $m_{0}=0$, massless Dirac cones
appear at the $X$ and $Y$ points [Fig.\ref{FigSpinConfig}(b)]. The term $%
m_{0}\tau _{x}$ is understood to simulate the effect of a gap opening due to
hybridization between the front and back surfaces in a thin film. The Dirac
cones at the $\Gamma $ and $M$ points are removed from the low-energy theory
when we take a large value of $m_{1}$.

The low-energy Dirac theory is extracted from (\ref{ThinFilm}) around the $X$
and $Y$ points as\beginABC\label{LowDirac}%
\begin{eqnarray}
H_{X} &=&(v_{1}k_{x}\sigma _{x}-v_{2}k_{y}\sigma _{y})\tau _{x}+m\tau _{x},
\\
H_{Y} &=&(v_{2}k_{x}\sigma _{x}-v_{1}k_{y}\sigma _{y})\tau _{x}+m\tau _{x},
\end{eqnarray}%
\endABC with the velocities $v_{1}$ and $v_{2}$ being given by%
\begin{equation}
v_{1}=\lambda _{1}-\lambda _{2},\quad v_{2}=\lambda _{1}+\lambda _{2}.
\end{equation}%
It is worthwhile to notice that this low-energy Hamiltonian agrees with the
one derived based on a first-principle calculation and the Dirac theory of
the TCI surface\cite{LiuFu}.

Our Hamiltonian is capable to simulate various models by controlling $%
m_{\ell}$. For instance, a massless Dirac cone emerges only at the $\Gamma $
point by setting $m_{0}=-2m_{1}$, as illustrated in Fig.\ref{FigBulkBand}%
(a3) and (b3). Similarly a massless Dirac cone emerges only at the $M$ point
by setting $m_{0}=2m_{1}$, as illustrated in Fig.\ref{FigBulkBand}(a4) and
(b4).

The thin film is an insulator, since a gap is given to all Dirac
points by the term $m_{0}\tau _{x}$. It is a topological insulator indexed
by the mirror-Chern number in the absence of the electric field ($E_{z}=0$).
It is a symmetry protected topological number.

As we derive in the Methods, the mirror-Chern charge may be calculated\cite%
{SPT} even for $E_{z}\neq 0$, and is given by%
\begin{equation}
C_{M}(E_{z})=\frac{1}{2}n_{x}n_{y}\frac{m}{\sqrt{m^{2}+E_{z}^{2}}}
\label{MirrorCME}
\end{equation}%
for each Dirac cone possessing the chirality $n_{x}n_{y}$ with $m$ given by (%
\ref{Az}). When $E_{z}=0$, it is reduced to%
\begin{equation}
C_{M}=\frac{1}{2}n_{x}n_{y}\text{sgn}(m).  \label{EachMC}
\end{equation}%
The total mirror-Chern number is quantized and given by 
\begin{equation}
C_{M}=\text{sgn}(m_{0}+2m_{1})+\text{sgn}(m_{0}-2m_{1})-2\text{sgn}(m_{0}).
\label{TotalMC}
\end{equation}%
We show the topological phase diagram in the $(m_{0},m_{1})$ plane in Fig.%
\ref{FigBulkBand}(e).

\textit{Nanoribbons:} With the tight-binding Hamiltonian at hand, we are
able to demonstrate the band structure of nanoribbons, as shown in Fig.\ref%
{FigBulkBand} for various values for parameters $m_{0}$ and $m_{1}$. We have
still set $E_{z}=0$. We take the direction of nanoribbon as $x$-axis. The
momentum component $k_{x}$ in the bulk band gives the momentum $k$ of
nanoribbon, while the momentum component $k_{y}$ is quantized. Accordingly,
the $X$ and $M$ points are projected to the same momentum $k=\pi $, while
the $Y$ and $\Gamma $ points are projected to $k=0$. The projected view of
the bulk band shown in Fig.\ref{FigBulkBand}(b) is the same as the band
structure of nanoribbon except for the edge states. Namely we can identify
the edge states by comparing the projected band structure of the bulk band
and the band structure of nanoribbon. The edge states are shown in magenta
in Fig.\ref{FigBulkBand}(c) and (d).

The bulk-edge correspondence works perfectly well. Indeed, gapless edge
modes emerge when a nanoribbon has a nonzero mirror-Chern number. It is
interesting that the gapless edge states emerge at $k=0$ for $C_{M}>0$ and
at $k=\pi $ for $C_{M}<0$. There are no edge states when the system is
trivial ($C_{M}=0$). There is exactly one to one correspondence between the
mirror-Chern number and the appearance of edge states in the band structure
of a nanoribbons.

\begin{figure*}[t]
\centerline{\includegraphics[width=0.9\textwidth]{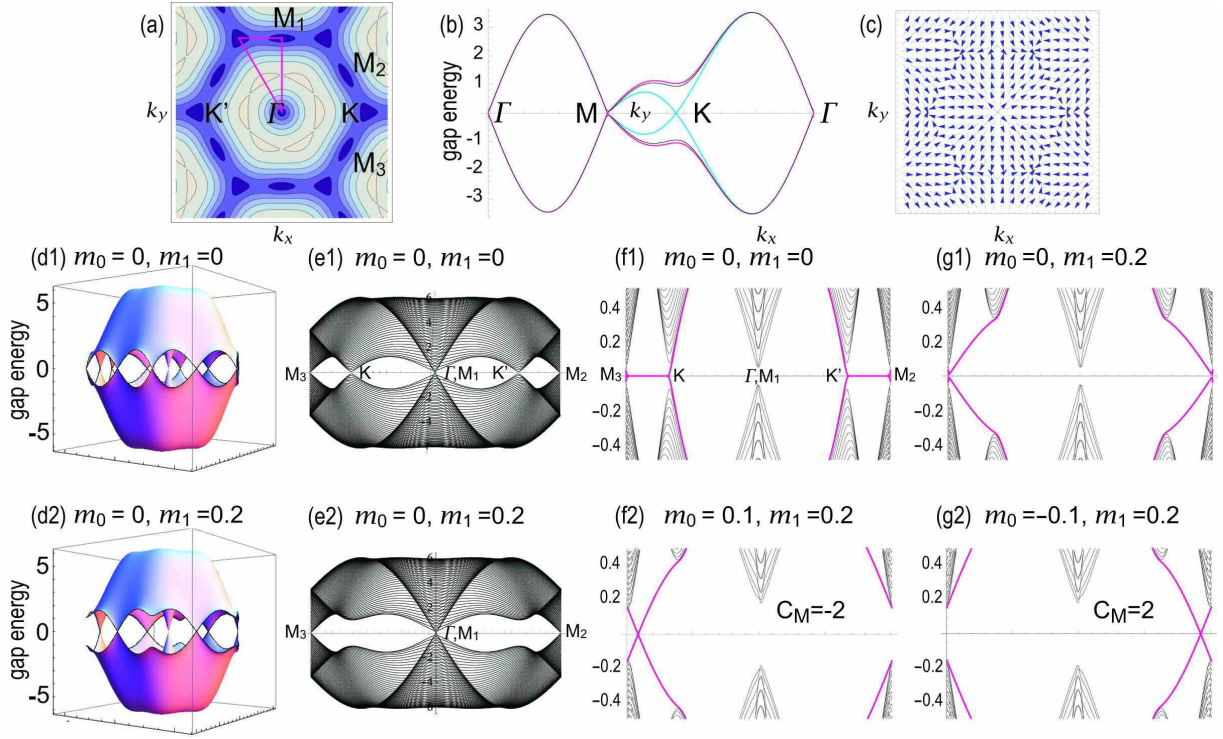}}
\caption{\textbf{Band structure and spin direction for triangular lattice.}
(a) We show a contour plot of the band structure of the Hamiltonian (\protect
\ref{HamilTotal}) on the triangular lattice. We have set $\protect\lambda %
_{1}=0$, $m_{1}=m_{2}=0$. The energy is lower in darker region. (b) We show
the band structure along the lines shown in red in (a). We have set $\protect%
\lambda _{1}=1$ in all three curves. Cyan curves are with $m_{1}=m_{2}=0$.
Magenta curves are with $m_{1}=0.2$, $m_{0}=0$. Black curves are with $%
m_{1}=0.2$, $m_{0}=0.1$. 
(c) We show the spin direction in the thin film with the $C_{6}$ symmetry for
the whole Brillouin zone.
(d) We present bird's eye views
and (e) projected views of the bulk band structure. (f),(g) We present the
band strucure of nanoribbons, where the gapless edge states are depicted in
the magenta curves. They emerge in topological insulators with nonzero
mirror-Chern numbers. We have take $\protect\lambda _{1}=1$ in all figures.
The values of $m_{0}$ and $m_{1}$ are indicated in each figure. The vertical
axis is the energy in unit of $\protect\lambda _{1} $ in all figures. The
horizontal axes are $-\protect\pi <k_{x}\leq \protect\pi $, $-\protect\pi %
<k_{y}\leq \protect\pi $ in (d), $-\protect\pi <k_{x}\leq \protect\pi $ in
(e), $-\protect\pi <k\leq \protect\pi $ in (f) and (g). }
\label{FigTriBand}
\end{figure*}

The nonzero mirror-Chern number indicates "quantum mirror Hall effects".
However it is a highly nontrivial problem to experimentally detect the
"mirror-Hall conductivity" since "mirror-Hall currents" convey neither
charge nor spin. On the other hand, there emerge $\left\vert
C_{M}\right\vert $ gapless states in the edges of a nanoribbon made of a
topological insulator with the mirror-Chern number $C_{M}$. Without the
electric field, these edge states transport merely the mirror charge $M$. Once
we apply external electric field parallel to the nanoribbon direction,
one edge state contributes one quantum unit to the electric conductance, as
we show in Fig.\ref{FigEz}(d1). Hence we are able to determine the absolute
value of the mirror-Chern number by measuring the conductance.

\textit{Electric field:} We now switch on the electric field $E_{z}$ between
the front and back surfaces to control the edge modes and the conductance in
nanoribbons [Fig.\ref{FigEz}(a)]. The mirror symmetry is broken by the
electric field as in (\ref{MirrorE}).

We show the band structure of a nanoribbon under the electric field $%
E_{z}$ in Fig.\ref{FigEz}(b) and (c). The edge states become gapped due to
the mixing of the right and left going edge states as a result of the
mirror-symmetry breaking.

The gapless edge mode transports the electric current. We have calculated
the conductance in the presence of $E_{z}$, which we show in Fig.\ref{FigEz}%
(d): See the Methods for derivation. The conductance near the Fermi energy
is $2$ for $E_{z}=0$ [Fig.\ref{FigEz}(d1)] since the edge states are doubly
degenerate. Once we turn on the electric field, the conductance falls to
zero since the edge states disappear due to the anticrossing [Fig.\ref{FigEz}%
(d2)]. Namely, it acts as a field-effect transistor\cite{LiuFu}. It is
possible to apply different electric fields $E_{z1}$ and $E_{z2}$ to the
right and left edge states [Fig.\ref{FigEz}(a)]. The conductance can be $0$, 
$1$ and $2$, which forms a multi-digit field-effect topological transistor
[Fig.\ref{FigEz}(d4)]. The conductance is quantized and topologically
protected.

\textbf{Triangular lattice with }$C_{6}$ \textbf{symmetry.} We proceed to
study the triangular lattice with the $C_{6}$ symmetry. Note that the
triangular lattice has the hexagonal symmetry. By substituting $N=3$ into
the Hamiltonian (\ref{Hamil}), and taking only the nearest neighbor sites ($%
\ell =1$), we obtain 
\begin{eqnarray}
A_{x} &=&-2\lambda _{1}\left[ \sin \frac{k_{x}}{2}\cos \frac{\sqrt{3}k_{y}}{2%
}+\sin k_{x}\right] , \\
A_{y} &=&-2\sqrt{3}\lambda _{1}\cos \frac{k_{x}}{2}\sin \frac{\sqrt{3}k_{y}}{%
2}.
\end{eqnarray}%
We show the band structure in Fig.\ref{FigTriBand}. There are six massless
Dirac cones, in which one Dirac cone resides at $\Gamma $, three Dirac cones
at $M$ points and two Dirac cones at $K$ and $K^{\prime }$
points. In the vicinity of each Dirac cone, we obtain the low-energy Dirac
theory 
\begin{equation}
A_{x}=v_{x}k_{x},\quad A_{y}=v_{y}k_{y},
\end{equation}%
with a set of velocities $(v_{x},v_{y})$ to be $(-3\lambda _{1},-3\lambda
_{1})$ for $\Gamma $, $(-\lambda _{1},3\lambda _{1})$ for $M_{1}$, $%
(3\lambda _{1}/2,3\lambda _{1}/2)$ for $K$ and $K^{\prime }$. The
chiralities of the Dirac cone at the $\Gamma $, $K$ and $K^{\prime }$ points
are identical, while three $M$ points have opposite chirality. It is
contrasted to the case of graphene, where the chiralities of $K$ and $%
K^{\prime }$ points are opposite.

We show the band structure of nanoribbons in Fig.\ref{FigTriBand}. It is
interesting that there exists a flat band in the region $-\pi\leq k\leq -\frac{2\pi }{3}$ and $\frac{2\pi }{3}\leq
k\leq \pi$: See Fig.\ref{FigTriBand}(f1) . The one is connecting
the $K$ and $M$ points, and the other is connecting $K^{\prime }$ and $M$
points.

In the similar manner to the square lattice, we introduce the mass term (\ref%
{Hamil-M}) to the Hamiltonian. The leading and the next leading terms are 
\begin{equation}
m=m_{0}+4m_{1}\sin \frac{k_{x}}{2}\left[ \cos \frac{\sqrt{3}k_{y}}{2}-\cos 
\frac{k_{y}}{2}\right] .  \label{TriMass}
\end{equation}%
By introducing the mass term, the Dirac cones at the $K$ and $K^{\prime }$
points become gapped. The resultant spectrum has four Dirac cones, in which
one Dirac cone resides at the $\Gamma $ point and the other three Dirac
cones at the $M$ points.

The total mirror-Chern number is given by 
\begin{equation}
C_{M}=\text{sgn}(m_{0}+3\sqrt{3}m_{1})+\text{sgn}(m_{0}-3\sqrt{3}m_{1})-2%
\text{sgn}(m_{0}).
\end{equation}%
This should be compared with the mirror-Chern number (\ref{TotalMC}) in the
square lattice with the $C_{4}$ symmetry. It follows that the phase diagram
is essentially given by the same one as Fig.\ref{FigBulkBand}(e).

We show the band structure of nanoribbons in Fig.\ref{FigTriBand} in the
presence of the mass term (\ref{TriMass}). The flat bands turn into the
dispersive edge modes. The position of the edge modes is between the $K$ ($%
K^{\prime }$) and $M_{2}=M_{3}$ points when $C_{M}<0$ ($C_{M}>0$). As in the
case of the square lattice, there is a perfect agreement between the
mirror-Chern number and the edge states of nanoribbons, as dictated by the
bulk-edge correspondence.

\section{Discussions}

The minimal tight-binding Hamiltonian of a TCI thin film is a four-band
model in order to take into account the spin and pseudospin (surface)
degrees of freedom. We have constructed such a model based on the symmetry
analysis. The prominent features are that gapless Dirac cones emerge at all
the high symmetric points and that we can provide them with gaps
phenomenologically at our disposal.

We have analyzed the square lattice with the $C_{4}$ symmetry and the triangular
lattice with the $C_{6}$ symmetry in details. The results may well describe
the thin films made of the [001] surface ($C_{4}$ symmetry) and the [111]
surface ($C_{6}$ symmetry) made of Pb$_{x}$Sn$_{1-x}$Te, by choosing the
mass parameters appropriately. According to experimental observations and a
first-principle calculation there are large gaps at the $\Gamma $ and $M$
points in the [001] surface\cite{Ando,Xu,Dz,LiuFu}. This is realized by
taking a large value of $m_{1}$ in our model. On the other hand there are
small gaps at the $X$ and $Y$ points, which is taken care of by introducing
a small value of $m_{0}$.

We may similarly discuss the square lattice with the $C_{2}$ symmetry with
the mass term being $m=m_{0}+m_{1}\cos k_{x}$. When $m_{0}=m_{1}$, there are
Dirac cones only $X$ and $M$ points, as is consistent with theoretical
results\cite{LiuFu,Safaei} on the [110] surface of Pb$_{x}$Sn$_{1-x}$Te. The
model with the $C_{3}$ symmetry is also constructed on the honeycomb
lattice. We find Dirac cones at the $K$ and $K^{\prime }$ points and two
degenerated Dirac cones at the $\Gamma $ point.

Our basic Hamiltonian consists of the SOI of the type $\mathbf{\sigma }\cdot 
\mathbf{d}_{ij}^{\ell }$ as in (\ref{Hamil}). We have made this choice since
it reproduces the low-energy Dirac theory\cite{LiuFu}. The same spectrum is obtained even if we take the
SOI of the Rashba type $\mathbf{\sigma }\times \mathbf{d}_{ij}^{\ell }$ as
in (\ref{HamilR}), although the low-energy Dirac theory now reads\beginABC%
\label{LowDiracR}%
\begin{eqnarray}
H_{X} &=&(v_{1}k_{y}\sigma _{x}+v_{2}k_{x}\sigma _{y})\tau _{x}+m\tau _{y},
\\
H_{Y} &=&(v_{2}k_{y}\sigma _{x}+v_{1}k_{x}\sigma _{y})\tau _{x}+m\tau _{y},
\end{eqnarray}%
\endABC and different from (\ref{LowDirac}). We predict that another TCI may
be found in future, where the Rashba-type Hamiltonian (\ref{HamilR}) plays
the basic role.

A TCI thin film may be used to design a nanodevice for topological
electronics. Edge states can be gapped by applying electric field
independently to the right and left edges. We have proposed a multi-digit
field-effect topological quantum transistor with the use of gapless edge
states of a TCI thin film nanoribbon. This could be a basic component of
future topological quantum devices.

\section{Methods}

In this section we explain the discrete rotational symmetry $C_{N}$. We also
describe how to calculate the mirror-Chern number and the conductance.

\textbf{Symmetry.} We have constructed the tight-binding Hamiltonian so that
it is invariant under the discrete rotation symmetry $C_{N}$ in addition to
the mirror symmetry (\ref{MirrorGener}). The generator of $C_{N}$ is 
\begin{equation}
C_{N}=R_{z}\exp \left[ -\frac{i\pi }{N}\sigma _{z}\right] ,
\end{equation}%
with the $\frac{2\pi }{N}$-rotation of the momentum%
\begin{equation}
R_{z}:\left( 
\begin{array}{c}
k_{x} \\ 
k_{y}%
\end{array}%
\right) \mapsto \left( 
\begin{array}{cc}
\cos \frac{2\pi }{N} & \sin \frac{2\pi }{N} \\ 
-\sin \frac{2\pi }{N} & \cos \frac{2\pi }{N}%
\end{array}%
\right) \left( 
\begin{array}{c}
k_{x} \\ 
k_{y}%
\end{array}%
\right) .
\end{equation}%
We note that the $C_{N}$ rotation rotates the direction of spin with $\pi /N$%
. The rotation angle is restricted to be $N=2,3,4,6$ due to the crystal
group of the lattice symmetry. They corresponds to the rectangular lattice
for $N=2$, the hexagonal lattice for $N=3$, the square lattice for $N=4$,
and the triangular lattice for $N=6$.

\textbf{Mirror-Chern number.} According to a general scheme\cite{SPT}, the
mirror-Chern charge is defined even for $E_{z}\neq 0$. With the use of the
Matsubara Green function, 
\begin{equation}
G\left( \mathbf{k}\right) =[i\omega -H\left( \mathbf{k}\right) ]^{-1},
\end{equation}%
with $i\omega $ referring to the Matsubara frequency ($\omega $: real), the
mirror-Chern charge is calculated by\cite{SPT} 
\begin{equation}
\mathcal{C}_{M}=\left( 2\pi \right) ^{-2}\int d^{2}k\int_{-\infty }^{\infty
}d\omega \,\Omega _{M}.  \label{ChernGamma}
\end{equation}%
Here, $\Omega _{M}=\frac{1}{6}\varepsilon _{\mu \upsilon \rho }$Tr$[G\Gamma
_{\mu }G\Gamma _{\nu }G\Gamma _{\rho }]$ and%
\begin{equation}
\Gamma _{x}=\frac{1}{2}\left\{ M,\partial _{x}G^{-1}\right\} ,\quad \Gamma
_{y}=\partial _{y}G^{-1},\quad \Gamma _{0}=\partial _{0}G^{-1},
\end{equation}%
with $M$ the mirror-symmetry generator (\ref{MirrorGener}). The result is
given by (\ref{MirrorCME}) for $E_{z}\neq 0$.

\textbf{Conductance.} In terms of single-particle Green's functions, the
low-bias conductance $\sigma (E)$ at the Fermi energy $E$ is given by\cite%
{Datta} 
\begin{equation}
\sigma (E)=(e^{2}/h)\text{Tr}[\Gamma _{\text{L}}(E)G_{\text{D}}^{\dag
}(E)\Gamma _{\text{R}}(E)G_{\text{D}}(E)],
\end{equation}%
where $\Gamma _{\text{R(L)}}(E)=i[\Sigma _{\text{R(L)}}(E)-\Sigma _{\text{%
R(L)}}^{\dag }(E)]$ with the self-energies $\Sigma _{\text{L}}(E)$ and $%
\Sigma _{\text{R}}(E)$, and%
\begin{equation}
G_{\text{D}}(E)=[E-H_{\text{D}}-\Sigma _{\text{L}}(E)-\Sigma _{\text{R}%
}(E)]^{-1},  \label{StepA}
\end{equation}%
with the Hamiltonian $H_{\text{D}}$ for the device region. The self-energy $%
\Sigma _{\text{L(R)}}(E)$ describes the effect of the electrode on the
electronic structure of the device, whose the real part results in a shift of
the device levels whereas the imaginary part provides a life time. It is to
be calculated numerically\cite{Sancho,Rojas,Nikolic,Li,EzawaAPL}.

\Section{Acknowledgements}

I am very much grateful to N. Nagaosa, L. Fu and T. H. Hsieh for many
helpful discussions on the subject. This work was supported in part by
Grants-in-Aid for Scientific Research from the Ministry of Education,
Science, Sports and Culture No. 22740196.

\Section{Additional information}

\textbf{Competing financial interests:} The author declares no competing
financial interests.

\textbf{Author contributions:} M. E. performed all calculations and made all
contribution to the preparation of this manuscript.

\end{document}